\documentclass[aps,preprint]{revtex4}%
\usepackage{amsfonts}%
\usepackage{amsmath}%
\usepackage{amssymb}%
\usepackage{epsfig}
\usepackage{subfigure}

\begin{document}

\title{RADIATIVE DECAY ENGINEERING BY TRIAXIAL NANOELLIPSOIDS}
\author{D.V. Guzatov}
\affiliation{P.N. Lebedev Physical
Institute, Russian Academy of Sciences, 53 Leninsky Prospect,
Moscow 119991, Russia}
\author{V.V. Klimov}
\email{vklim@sci.lebedev.ru}
\affiliation{P.N. Lebedev Physical
Institute, Russian Academy of Sciences, 53 Leninsky Prospect,
Moscow 119991, Russia}


\begin{abstract}

 Radiative decay rates of an atom placed near triaxial
nanoellipsoid are investigated in long wavelength approximation.
Analytical results are obtained in general case. It is shown that
triaxial ellipsoid can be used for efficient control of decay rate
of an atom, molecule or quantum dot. For example decay rate near
silver ellipsoid can be enhanced by \textit{5} orders of
magnitude. It is also shown, that triaxial nanoellipsoid can be
used for simultaneous efficient control of absorption and emission
rates of fluorophores.
\end{abstract}

%\volumeyear{}
%\volumenumber{}
%\issuenumber{}
%\eid{}

\date[]{September 15, 2004}
\startpage{1}
\endpage{2}
\maketitle

 Because of development of nanotechnology, investigation
of interaction of single atoms, molecules and quantum dots with
optical fields in presence of nanobody becomes more and more
topical. Special attention must be given here on spontaneous decay
of atoms placed near nanobodies. Spontaneous emission is, above
all, a source of light and its efficient control allows increase
efficacy of light sources \cite{ref1}. Spontaneous emission of
single molecule can be used as a source of nanolocalized light
\cite{ref2}, with help of such kind of sources, it is possible to
examine nanometer-sized objects. On the other hand, decay rate is
directly measured magnitude at fluorescent detection and
identification of single molecules by scanning microscopes
\cite{ref3}. Influence of nanoparticles on fluorescence of single
molecule can be used for detection and identification of small
amount of molecules and for sequencing of DNA structure, without
using additional fluorescent markers \cite{ref4}. It is highly
important to say, that if it is be possible to match fluorescent
properties of nanoparticle with properties of detected molecule,
then it will be possible to provide high selectivity of detection.
On the other hand, nanotechologies allow creating complex
nanosystems, consisted of radiating nanoparticle (finite
nanocylinder of semiconductor) and of nanoresonator (gold
nanosphere) \cite{ref5}. Besides this, the system of nanobody +
nanoradiator (QD) can be used as amplifier of surface plasmons by
stimulated emission of radiation (SPASER) \cite{ref6}. And of
course, it is obvious, that radiative decay rate is an important
parameter, which influence on atoms' motion in optical fields of
nanometer scale \cite{ref7}. In all these cases, the main task is
correct description of process of spontaneous decays near the
nanoparticles of different shapes.

\begin{figure}
\centering \epsfig{figure=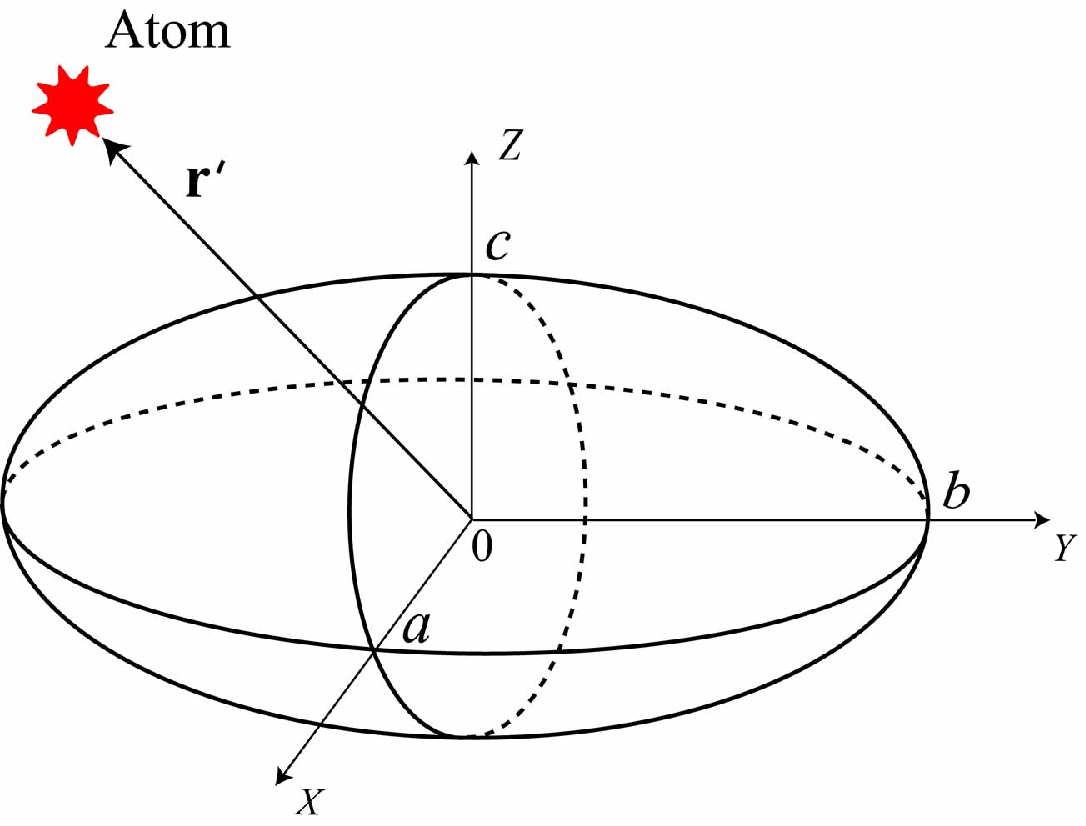,width=.5 \textwidth}
\renewcommand{\figurename}{Fig.}
\caption{Geometry of the problem.} \label{fig1}
\end{figure}

The aim of this work is an investigation of processes of
spontaneous decay of excited atom (molecule or quantum dot) in the
presence of ellipsoid-shaped nanobody with arbitrary ratio of
principal semiaxes. The geometry of the considered system is
presented on Fig.~\ref{fig1}. Such general shape of the nanobody
has not only academic interest, but also a practical one, because
of possibility of control three dimensional parameters (or two
dimensionless ones) allows describe nanoparticles of different
shapes (nanospheres, nanoneedles, nanodisks, nanowires, and
others) and it leads to wide possibilities in controlling of
spontaneous emission decay rates of an atom, placed near of such
nanobody, and in construction on this base of efficient
nanosensors. Besides this, optical properties of nanoellipsoid and
related local fields will suffer substantial change at different
variations of nanoellipsoid's shape.

In case of nanoparticles, relative radiative decay rate of an atom
is defined by its dipole momentum only (see, for example,
\cite{ref8})

\begin{equation}
\left( \frac{\gamma }{\gamma _{0}}\right)
^{radiative}=\frac{\left\vert \mathbf{d}_{total}\right\vert
^{2}}{\left\vert \mathbf{d}_{0}\right\vert ^{2} },  \label{eq1}
\end{equation}

where \textbf{d}$_{total}$\ is total dipole momentum of the atom +
nanobody system; \textbf{d}$_{0}$ is dipole momentum of an atom's
transition. Therefore, in this case, our task is to find total
dipole momentum of the considered system in quasistatic approach.

To find the total dipole momentum, we need to solve a quasistatic problem
with dipole source

\begin{equation}
rot\mathbf{E}=0,\quad div\mathbf{D}=4\pi \rho ,  \label{eq2}
\end{equation}

where density of the charge is defined by the standard expression:

\begin{equation}
\rho =e^{-i\omega t}\left( \mathbf{d}_{0}\cdot \nabla ^{\prime }\right)
\delta ^{\left( 3\right) }\left( \mathbf{r}-\mathbf{r}^{\prime }\right) .
\label{eq3}
\end{equation}

Here $\delta ^{\left( 3\right) }$ is the spatial Dirac's
delta-function; \textbf{r}$^{\prime }$ is vector of coordinate of
the atom; ${\nabla }^{\prime }$ means gradient over the atom's
coordinates. Hereafter we will omit the time dependence of the
field. It is convenient to rewrite system (\ref{eq2}) in the form
of integral equation:

\begin{equation}
\mathbf{E}\left( \mathbf{r}\right) =\mathbf{E}_{0}\left( \mathbf{r}\right) -%
\frac{\epsilon -1}{4\pi }{\int\limits_{V}}d\mathbf{r}^{\prime
}\left( \mathbf{E}\left( \mathbf{r}^{\prime }\right) \cdot \nabla
^{\prime }\right) \nabla ^{\prime }\frac{1}{\left\vert
\mathbf{r}-\mathbf{r}^{\prime }\right\vert }.  \label{eq4}
\end{equation}

Here \textbf{E}$_{0}\left( \mathbf{r}\right) $ is dipole field in the
absence of nanobody; and integration is taking over the nanobody's (the
ellipsoid) volume $V$. Permittivity of the ellipsoid is denoted by $\epsilon
$; and permittivity of the surrounding medium we put equal to unit.

As it is known, induced dipole momentum of a body is defined by
following formula

\begin{equation}
\delta \mathbf{d}=\frac{\epsilon -1}{4\pi }{\int\limits_{V}}d\mathbf{rE}%
\left( \mathbf{r}\right) ,  \label{eq5}
\end{equation}

and we need only to calculate integral over the nanobody's volume.
To calculate it, we can use expression (\ref{eq4}), and to write

\begin{equation}
{\int\limits_{V}}d\mathbf{rE}\left( \mathbf{r}\right) ={\int\limits_{V}}d%
\mathbf{rE}_{0}\left( \mathbf{r}\right) -\frac{\epsilon -1}{4\pi }{%
\int\limits_{V}}d\mathbf{r}^{\prime }\left( \mathbf{E}\left( \mathbf{r}%
^{\prime }\right) \cdot \nabla ^{\prime }\right) \nabla ^{\prime }{%
\int\limits_{V}}d\mathbf{r}\frac{1}{\left\vert
\mathbf{r}-\mathbf{r}^{\prime }\right\vert }. \label{eq6}
\end{equation}

Some amazing property of ellipsoid is that the last integral in
(\ref{eq6}) depends on position of point \textbf{r}$^{\prime }$
inside the ellipsoid in quadratic manner, and in that way we will
come to algebraic system of equation for induced dipole momentum

\begin{equation}
\delta \mathbf{d}+ \frac{\epsilon -1}{4\pi }\left( \delta \mathbf{d\cdot }%
\nabla ^{\prime }\right) \nabla ^{\prime }{\int\limits_{V}}\frac{d\mathbf{r}%
}{\left\vert \mathbf{r}-\mathbf{r}^{\prime }\right\vert }=\frac{\epsilon -1}{%
4\pi }{\int\limits_{V}}d\mathbf{rE}_{0}\left( \mathbf{r}\right) ,
\label{eq7}
\end{equation}

which has the next solution

\begin{equation}
\delta \mathbf{d=}\widehat{\mathbf{\alpha }}\cdot \frac{1}{V}{\int\limits_{V}%
}d\mathbf{rE}_{0}\left( \mathbf{r}\right) ,  \label{eq8}
\end{equation}

where $V={\frac{{4}}{{3}}}\pi abc$ is ellipsoid's volume; and $\widehat{%
\mathbf{\alpha }}$ is the tensor of ellipsoid's polarizability

\begin{equation}
\widehat{\mathbf{\alpha }}=\frac{\epsilon -1}{4\pi }V\left( \mathbf{I}+\frac{%
\epsilon -1}{4\pi }\nabla ^{\prime }\nabla ^{\prime }{\int\limits_{V}}\frac{d%
\mathbf{r}}{\left\vert \mathbf{r}-\mathbf{r}^{\prime }\right\vert
}\right) ^{-1}.  \label{eq9}
\end{equation}

If the coordinate axes coincide with the principal axes, the
polarizability has well known form \cite{ref9}

\begin{equation}
\widehat{\mathbf{\alpha }}=\left(
\begin{array}{ccc}
\alpha _{xx} & 0 & 0 \\
0 & \alpha _{yy} & 0 \\
0 & 0 & \alpha _{zz}%
\end{array}%
\right) ,  \label{eq10}
\end{equation}

where

\begin{equation}
\alpha _{xx}=\frac{1}{4\pi }\left( 1-\epsilon _{xx}\right) \left( \frac{%
\epsilon -1}{\epsilon -\epsilon _{xx}}\right) ,\quad \epsilon
_{xx}=1-\left( \frac{1}{2}abcI_{a}\right) ^{-1},  \label{eq11}
\end{equation}

\begin{equation}
I_{a}={\int\limits_{0}^{\infty }}\frac{du}{\left( a^{2}+u\right)
R\left( u\right) },\quad R\left( u\right) =\left[ \left(
a^{2}+u\right) \left( b^{2}+u\right) \left( c^{2}+u\right) \right]
^{1/2}.  \label{eq12}
\end{equation}

Other components of tensor (\ref{eq10}) can be obtained by cyclic
permutation of all parameters and indexes.

One should note, that expressions (\ref{eq12}) can be expressed
through complete and incomplete elliptical integrals in general
case. It can be expressed through Legendre's functions of first
and second type or through elementary functions in case of
degeneration of triaxial ellipsoid into spheroid or sphere
\cite{ref10}-\cite{ref11}.

In this way, presented above expression for the induced dipole momentum in
an arbitrary external field can be expressed by the next formula

\begin{equation}
\delta \mathbf{d}=\widehat{\mathbf{\alpha }}\cdot \left\langle \mathbf{E}%
_{0}\right\rangle ,  \label{eq13}
\end{equation}

where $\left\langle \mathbf{E}_{0}\right\rangle $ is the field of
external source averaged over ellipsoid's volume. This expression
is a key mark one, because of possibility of solving of the
considered problem just by averaging over the nanobody's volume.
Expression (\ref{eq13}) can be also used as an estimate one for
dipole momentum of nanoparticle of other geometry (like nanocube,
finite nanocylinder, and others). Of course, we can use it in that
case only, when polarizability of nanoparticle is known.

In our case, external field is the electric field from dipole situated at point \textbf{r}$%
^{\prime }$

\begin{equation}
\mathbf{E}_{0}\left( \mathbf{r}\right) =\nabla \left(
\mathbf{d}_{0}\cdot \bigtriangledown \right) \frac{1}{\left\vert
\mathbf{r}-\mathbf{r}^{\prime }\right\vert }.
\label{eq14}
\end{equation}

An average over the ellipsoid's volume given by (\ref{eq14}) can
be expressed through one-dimensional integrals \cite{ref12}. As a
result we have

\begin{equation}
\left\langle \mathbf{E}_{0}\right\rangle =\nabla \left(
\mathbf{d}_{0}\cdot \bigtriangledown \right) J, \label{eq15}
\end{equation}

where

\begin{equation}
J=\frac{3}{4}\left\{ I\left( \zeta \right) -x^{2}I_{a}\left( \zeta
\right) -y^{2}I_{b}\left( \zeta \right) -z^{2}I_{c}\left( \zeta
\right) \right\} , \label{eq16}
\end{equation}

\begin{eqnarray}
I\left( \zeta \right) =\int\limits_{\zeta }^{\infty
}\frac{du}{R\left( u\right) },\quad \quad \quad \quad \quad
I_{a}\left( \zeta \right) =\int\limits_{\zeta }^{\infty
}\frac{du}{\left( a^{2}+u\right) R\left( u\right) }, \nonumber \\
I_{b}\left( \zeta \right) =\int\limits_{\zeta }^{\infty
}\frac{du}{\left( b^{2}+u\right) R\left( u\right) },\quad
I_{c}\left( \zeta \right) =\int\limits_{\zeta }^{\infty
}\frac{du}{\left( c^{2}+u\right) R\left(
u\right) },%
\label{eq17}
\end{eqnarray}

and where function $R\left( u\right) $ is defined by formula
(\ref{eq12}). Hereafter we will omit primes subscript in atom's
coordinates.

In case of atom, located outside the ellipsoid, the point $\zeta
$\ is a
positive root of the cubic equation: $\frac{x^{2}}{a^{2}+\zeta }+%
\frac{y^{2}}{b^{2}+\zeta }+\frac{z^{2}}{c^{2}+\zeta }=1$;
and it follows in this case, that $\zeta $\ is the function of
coordinates of atom's position. In the case, when atom is situated
inside the ellipsoid, we must put $\zeta =0$ and it means that
spontaneous emission decay rate does not depend on atom's
position.

Thus, spontaneous emission decay rate of an single atom located
near triaxial nanoellipsoid, and when atom has arbitrary oriented
dipole momentum, has the following form

\begin{equation}
\left( \frac{\gamma }{\gamma _{0}}\right)
^{radiative}=\frac{\left\vert
\mathbf{d}_{total}\right\vert ^{2}}{\left\vert \mathbf{d}_{0}\right\vert ^{2}%
},\quad \delta \mathbf{d}=\widehat{\mathbf{\alpha }}\cdot \nabla
\left( \mathbf{d}_{0}\cdot \nabla \right) J, \label{eq18}
\end{equation}

where $J$ is defined by formula (\ref{eq16}).

Expression (\ref{eq18}) is a fundamental result, and we can use it
to obtain well known expressions for spontaneous emission decay
rate of an atom placed near nanospheroid \cite{ref10}-\cite{ref11}
and near nanocylinder \cite{ref8}.

In some particular cases, expression for relative decay rate
(\ref{eq18}) can be substantially simplified. For example, when
atomic dipole is situated on \textit{z}-axis of the Cartesian
system at point $z$ and it oriented along \textit{x}-axis we have

\begin{equation}
\left( \frac{\gamma }{\gamma _{0}}\right) ^{radiative}=\left\vert 1-\frac{1}{%
2}abc\left( 1-\epsilon _{xx}\right) \left( \frac{\epsilon
-1}{\epsilon -\epsilon _{xx}}\right) I_{a}\left(
z^{2}-c^{2}\right) \right\vert ^{2}. \label{eq19}
\end{equation}

In that case when atom is situated on \textit{z}-axis at point $z$
and its dipole momentum is oriented along \textit{z}-axis also we
have

\begin{eqnarray}
\left( \frac{\gamma }{\gamma _{0}}\right) ^{radiative} &=&\left\vert 1+\frac{%
1}{2}abc\left( 1-\epsilon _{zz}\right) \left( \frac{\epsilon
-1}{\epsilon -\epsilon _{zz}}\right) \times \right.   \label{eq20}
\\ &&\left. \times \left\{ \frac{2}{\sqrt{\left(
z^{2}+a^{2}-c^{2}\right) \left( z^{2}+b^{2}-c^{2}\right)
}}-I_{b}\left( z^{2}-c^{2}\right) \right\} \right\vert ^{2}.
\nonumber
\end{eqnarray}

It is important to note, that founded expressions do not have
sense (became infinite large) at some (negative) values of
permittivity. This phenomenon corresponds to cases of plasmon
resonance excitation. If we take into account higher terms of
expansion over inverse wavelength (radiative corrections) than
this divergence disappeares \cite{ref13}, but plasmon resonance
remains.

Highly important moment is that for arbitrary value of $\epsilon \left( {%
\lambda }\right) $, i.e. for arbitrary material at arbitrary
frequency, it is possible to find a whole series of ellipsoids of
different shapes having plasmon resonances. This variety is a base
for different effects.

On the other hand, slightly changing of the ellipsoid's shape with
settled value of permittivity is a reason of radical changing in
spontaneous emission decay rate.

\begin{figure}
\centering \subfigure[$b/c=0.6$ and %
$a/c=0.105$]{\epsfig{figure=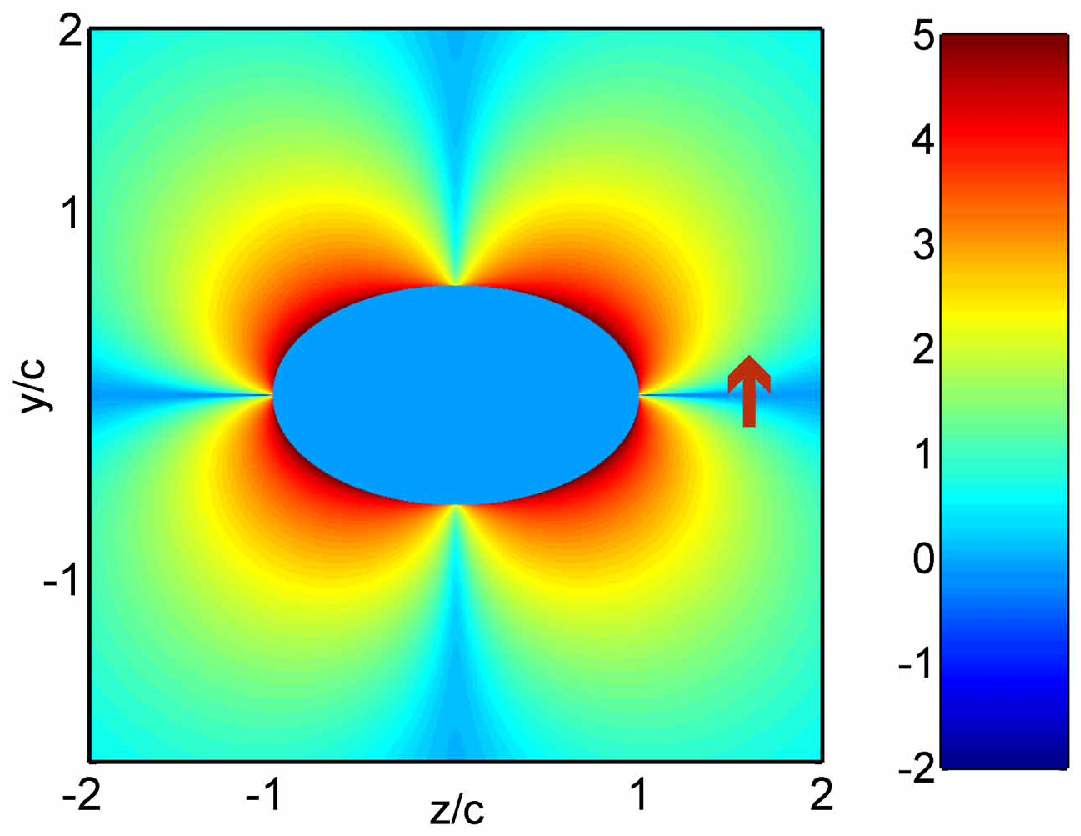,width=.6 \textwidth}}\quad
\subfigure[$b/c=0.6$ and %
$a/c=0.046$]{\epsfig{figure=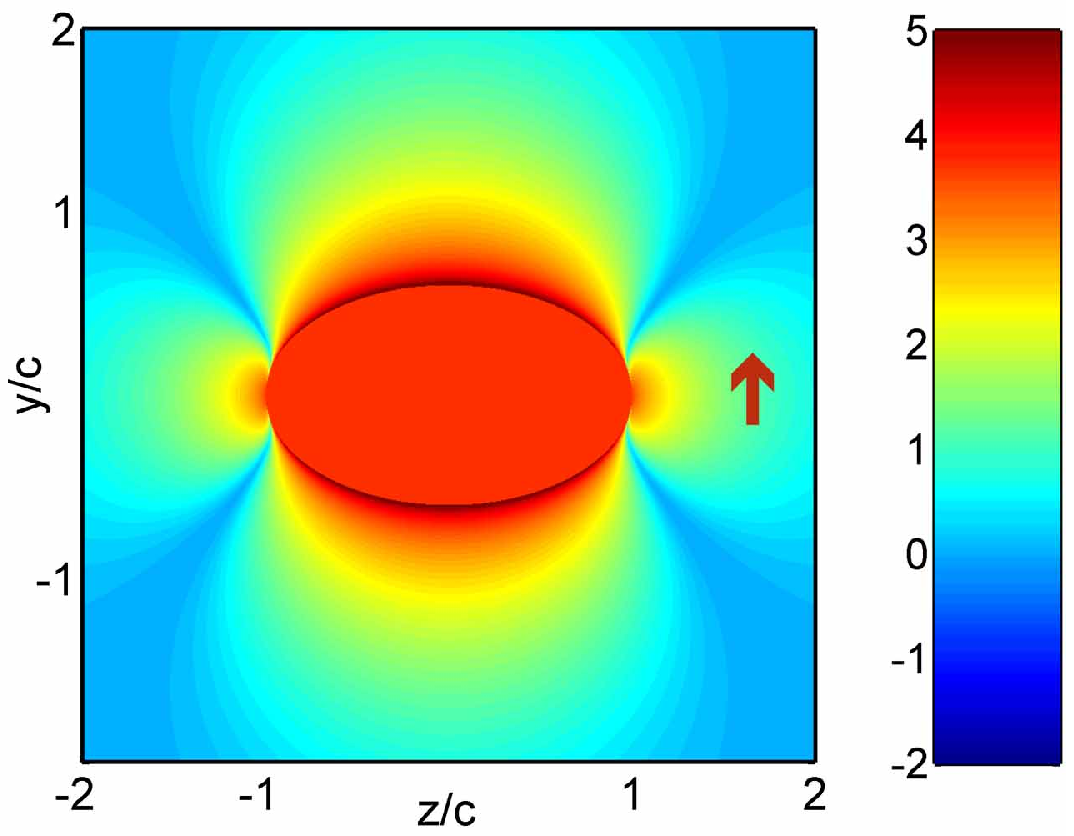,width=.6 \textwidth}}
\renewcommand{\figurename}{Fig.}
\caption{Relative radiative decay rate of an atomic dipole with
orientation of the moment along $y$-axis and placed near silver
nanoellipsoid ($\epsilon =-15.37+i0.231$ ($\lambda =632.8$ nm)) as
a function of atom's position in $x=0$ plane. Red arrow indicates
orientation of dipole momentum. Colorbar is in logarithmic scale.}
\label{fig2}
\end{figure}

\begin{figure}
\centering \subfigure[$b/c=0.6$ and %
$a/c=0.105$]{\epsfig{figure=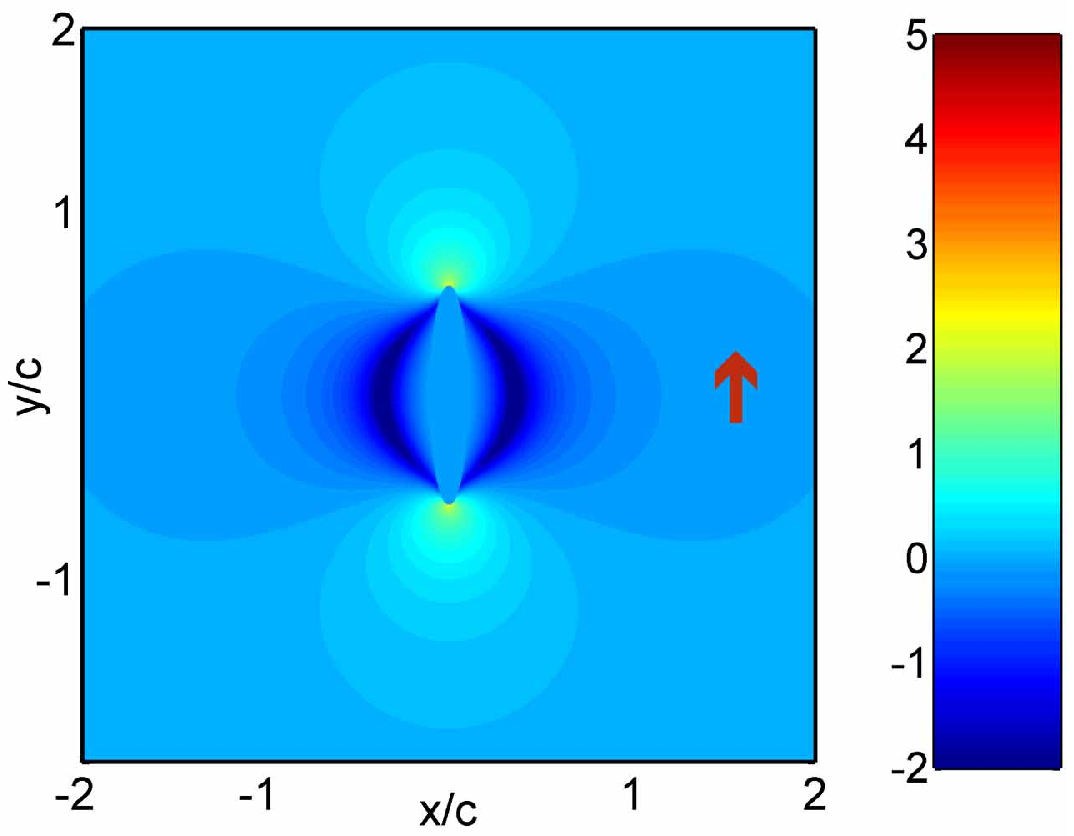,width=.6 \textwidth}}\quad
\subfigure[$b/c=0.6$ and %
$a/c=0.046$]{\epsfig{figure=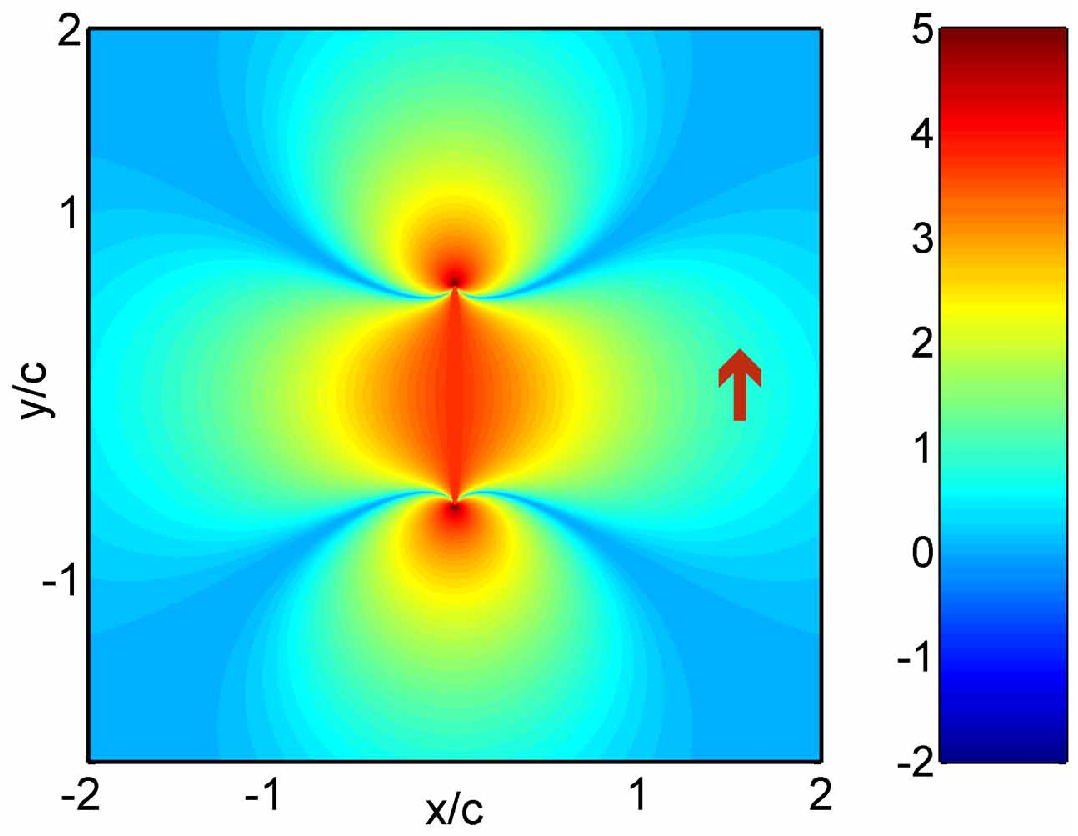,width=.6 \textwidth}}
\renewcommand{\figurename}{Fig.}
\caption{Relative radiative decay rate of an atomic dipole with
orientation of the moment along $y$-axis and placed near silver
nanoellipsoid ($\epsilon =-15.37+i0.231$ ($\lambda =632.8$ nm)) as
a function of atom's position in $z=0$ plane. Red arrow indicates
orientation of dipole momentum. Colorbar is in logarithmic scale.}
\label{fig3}
\end{figure}

On Figures ~\ref{fig2} and ~\ref{fig3} dependences of spontaneous
emission decay rate of an atom placed near silver ($\epsilon
=-15.37+i0.231$ ($\lambda =632.8$ nm)) nanoellipsoid on atom
position are presented. From Fig.~\ref{fig2}(a) ($b/c=0.6$, $%
a/c=0.105$) it is well seen substantial decrease of the radiative
decay rate inside of ellipsoid (where decay rate is constant) and
four regions of substantial enhancement of radiative decay exist
near ellipsoid's surface. Slightly changing of the ellipsoids
shape ($b/c=0.6$, $a/c=0.046$) we have another plasmon
excitation, as it presented on Fig.~\ref{fig2}(b). In this case,
in contrary to spontaneous emission process presented by
Fig.~\ref{fig2}(a), there is substantial acceleration of the
radiative decay rate of an atom placed inside the ellipsoid.

Analogous picture takes place in case of $z=0$ cross section of spontaneous
emission decay rate, as it presented on Fig.~\ref{fig3}. In that case, on Fig.~\ref{fig3}(a) ($%
b/c=0.6$, $a/c=0.105$) near the ellipsoid regions of substantial
prohibition of radiative decay occurs (from the left and from the
rights sides of ellipsoid's surface). In case of decreasing of
semi-axis $a$, in approximately down to two times ($b/c=0.6$,
$a/c=0.046$), the situation is changes to opposite one, as it
clearly presented on Fig.~\ref{fig3}(b). And as it was mentioned
above, instead of delay of spontaneous decays, now we have
substantial acceleration.

Thus, spontaneous emission decay rate near triaxial ellipsoids of
arbitrary shape depends on geometry of the ellipsoid and of
spatial localization of an atom in highly nontrivial way. In some
regions, there is substantial acceleration of transitions, but in
other regions substantial inhibition can occur. Analogously, in
the case of incidence of plain electromagnetic wave (i.e. dipole
source situated at infinity), the field distribution near triaxial
nanoellipsoid also has rather complicated structure, where regions
of substantial enhancement and substantial inhibition exist. All
above mentioned means, that triaxial ellipsoids can be used for
effective control of local plasmon fields and therefore, it can
replace more complicated constructions of simple nanoparticles
\cite{ref14}.

Of course, in case of strong acceleration of spontaneous emission decay
rates, the perturbation theory will be unusable for strong dipole
transitions. In this case, Rabi's oscillations can occur, but theory of such
processes is also can be developed on the base of the method presented in
our paper.

So far, we considered process of spontaneous emission, i.e. we
have assumed that an atom is in some excited state. But the
triaxial ellipsoid can be used for efficient excitation of the
atoms and molecules having their absorption bandwidth in resonance
with plasmon oscillations of the ellipsoid. Moreover, triaxial
ellipsoids (in contrast to spheroids and spheres) can be used for
\textit{simultaneous}\ amplification (or inhibition) of absorption
and emission. It is due to the fact that for every kind of
fluorophore, there such ellipsoid exists for which one of the
resonant frequencies lies in absorption bandwidth and the other
one lies in emission bandwidth. For example, if we consider
ellipsoid made of silver with the next aspect ratios $a/c=0.273$ and $b/c=0.447$, then it has plasmon resonances at $%
355$ nm and $465$ nm and these two lines correspond to frequency domains of
absorption and emission bandwidths of $Hoechst$ 33342 fluorophore \cite{ref15}%
. Analogously, silver ellipsoid with aspect ratios $a/c=0.357$ and
$b/c=0.414$ has plasmon resonances at $346$ nm and $445$ nm and
these two lines corresponds to absorption and emission bandwidth
of $Alexa$ 532 fluorophore \cite{ref15}. In its turn, the
simultaneous amplification of absorption and emission rates leads
to substantial increase of fluorescent intensity and also leads to
possibility of detection and observing of molecules with higher
temporal resolution. Discovered effect allows investigate that
fluorophores, which is uninteresting without using of the triaxial
nanoellipsoid (or other nanoparticles of complicated shapes).

In conclusion, analytical expressions for spontaneous emission
decay rate of a single atom placed near triaxial nanoellipsoid are
founded. Obtained results show, that for metallic ellipsoids there
is inhibition enhancement in several orders of magnitude of
spontaneous decay are possible for an atom (or a molecule) and it
is strongly depends on atom's location near the ellipsoid. It was
shown, that triaxial ellipsoids can be used for simultaneous
control of absorption and emission processes of light by single
molecule.

\begin{acknowledgements}
We would like to acknowledge of Russian Foundation of Basic
Research (grant 04-02-16211-a) and U.S. Civilian Research and
Development Foundation (grant RP2-2494-MO-03) for partial
financial support of this investigation.
\end{acknowledgements}

\newpage

\newpage

\newpage

\newpage

\end{document}